\documentclass[aps,prc,preprint,groupedaddress,preprintnumbers]{revtex4-1}
\usepackage{natbib}
\usepackage{amsmath}
\usepackage{graphicx}
\usepackage{hyperref}
\usepackage{slashbox}
\usepackage{longtable}
\usepackage{multirow}
\newcommand{\Pom}{I$\!$P}                % gives pomeron symbol
\newcommand{\physrep}{Physics Reports}
\begin{document}
\preprint{\vbox{\hbox{ JLAB-THY-12-1646}}}
% Use the \preprint command to place your local institutional report
% number in the upper righthand corner of the title page in preprint mode.
% Multiple \preprint commands are allowed.
% Use the 'preprintnumbers' class option to override journal defaults
% to display numbers if necessary
%\preprint{}

%Title of paper
\title{A Regge Model for Nucleon-Nucleon Spin Dependent Amplitudes}

% repeat the \author .. \affiliation  etc. as needed
% \email, \thanks, \homepage, \altaffiliation all apply to the current
% author. Explanatory text should go in the []'s, actual e-mail
% address or url should go in the {}'s for \email and \homepage.
% Please use the appropriate macro foreach each type of information

% \affiliation command applies to all authors since the last
% \affiliation command. The \affiliation command should follow the
% other information
% \affiliation can be followed by \email, \homepage, \thanks as well.
\author{William P. Ford$^{(1)}$}
\email[]{wpford@jlab.org}
\author{J. W. Van Orden$^{(1,2)}$}
\email[]{vanorden@jlab.org}
%\homepage[]{Your web page}
%\thanks{}
%\altaffiliation{}
\affiliation{\small \sl (1) Department of Physics, Old Dominion University, Norfolk, VA
23529\\and\\ (2) Jefferson Lab\footnote{Notice: Authored by Jefferson Science Associates, LLC under U.S. DOE Contract No. DE-AC05-06OR23177. 
The U.S. Government retains a non-exclusive, paid-up, irrevocable, world-wide license to publish or reproduce this manuscript for U.S. Government purposes}, 
12000 Jefferson Avenue, Newport News, VA 23606}
\date{\today}
\begin{abstract}
There are currently no models readily available that provide nucleon-nucleon spin dependent scattering amplitudes at high energies ($s \geq 6$  $ GeV^2$). 
This work aims to provide a model for calculating these high energy scattering amplitudes. 
The foundation of the model is Regge theory since it allows for a relativistic description and full spin dependence. 
We present our parameterization of the amplitudes, and show comparisons of our solutions to the data set we have collected. 
Overall the model works as intended, and provides an adequate description of the scattering amplitudes.
\end{abstract}
\maketitle
\section{Introduction}
We present a model for calculating elastic, spin-dependent scattering amplitudes for the nucleon-nucleon system. 
While much work has been applied to this topic over the years there is no analysis available for both proton-proton and proton-neutron 
in the mid to high energy range,  Mandelstam $s>6$ GeV$^2$. 
Thus far the most complete, highest energy, and readily available work is the SAID \cite{SAIDpaper,SAIDdata} 
analysis which provides the proton-neutron amplitudes to $s \approx 6$ GeV$^2$ and the proton-proton amplitudes up to $s \approx 9.8$ GeV$^2$. 
Our goal is to calculate the amplitudes at higher energies. In order to accurately describe the nucleon-nucleon system at these energies we require 
a fully relativistic, spin dependent model. Furthermore, due to the scarcity of data, particularly in the proton-neutron case, we require a model 
which will provide some confidence in extrapolating the results to higher energies. 

Our primary motivation in building this model is to utilize the amplitudes to describe the final state interactions of deuteron electrodisintegration. 
It has been shown that a complete description of the final state interactions is necessary in order to accurately describe this process \cite{JVO_2008_newcalc,JVO_2009_tar_pol,JVO_2009_ejec_pol}. 
Currently the final state interactions are given by the SAID amplitudes, however, the kinematics at Jlab, where experiments have been performed, 
allow for final state nucleons with energies greater than can be described by SAID. 

Our approach is to parameterize the helicity amplitudes in terms of Regge poles or exchanges \cite{regge1959,martin,perl,collins,irving_Regge_phenom}.
Regge theory has had great phenomenological success over the years. The theory is fully relativistic and allows for a complete spin-dependent description. 
Regge theory is also ideal for us to use since it readily scales to higher energies. Furthermore, Regge theory has been utilized in the past to model 
proton-neutron scattering at mid-range energies with good results \cite{irvingNN}. Overall Regge theory provides us a systematic method of parameterizing 
the scattering amplitudes, while meeting all the criteria of the model.

The fundamental idea of Regge theory is to study the analytic behavior of the amplitudes, when one allows the angular momentum $J$ to be continuous and complex. 
While the analysis is rigorous\cite{regge1959} for non-relativistic scattering, the relativistic case is based on a series of assumptions. 
However, it is in the relativistic case, by exploiting crossing symmetry, that Regge theory is useful as a parameterization method. 
Performing the Regge analysis in the $t$-channel center of momentum frame, the large $s$, small $t$ approximation to the amplitudes in the $s$-channel 
center of momentum frame is obtained \cite{martin,perl,collins}. 

While Regge theory excels at high energies, at lower energies it becomes more difficult to implement, as more and more Regge exchanges can contribute. 
Because of this feature, however, it naturally lends itself as a method for extrapolating to higher energies, 
since as one increases in energy the low energy poles are suppressed. 
In our approach we are able to fit to the low energy nucleon-nucleon data and extrapolate our results to higher energy regions where data is unavailable.

Section \ref{sec:theoretical framework} discusses our method of parameterizing the helicity amplitudes in terms of Regge poles. Then in Section \ref{sec:results} 
we present comparisons between the data and our results for both a polarized and unpolarized fit solution. 
We conclude in Section \ref{sec:summary} with a summary and outlook. 

\section{Theoretical Framework}
\label{sec:theoretical framework}
All observables in the nucleon-nucleon system can be described by five independent helicity amplitudes \cite{Bystricky}. These amplitudes are given as,
\begin{align} \label{eq:amplitudes(abcde)}
a &= \phi_1 = \langle ++ |T| ++ \rangle \nonumber\\
b &= \phi_5 = \langle ++ |T| +- \rangle \nonumber\\
c &= \phi_3 = \langle +- |T| +- \rangle \\
d &= \phi_2 = \langle ++ |T| -- \rangle \nonumber\\
e &= \phi_4 = \langle +- |T| -+ \rangle.\nonumber
\end{align}

\begin{figure}
    \includegraphics[width=8.6cm]{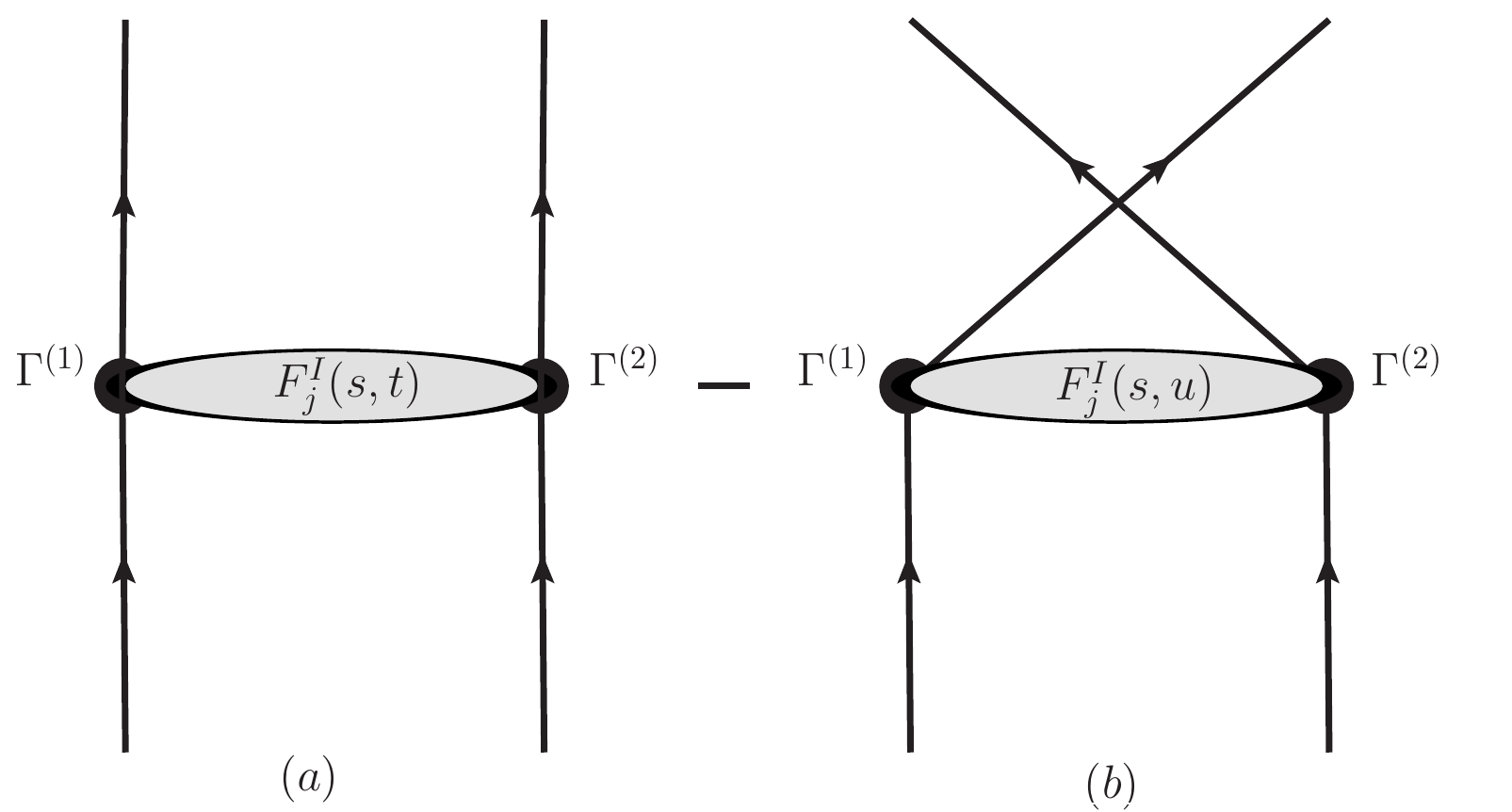} 
    \caption{Pictorial represention of the helicity amplitudes in terms of the Fermi invariants. $\Gamma$ represents the various gamma matrices which contribute to this process.}
    \label{fig:stat_wFI}
\end{figure}

In order to calculate the amplitudes we look to Regge theory to provide us with a parameterization method. 
Applying Regge theory to the nucleon-nucleon system presents some challenges, primarily due to the inclusion of spin. 
The Regge analysis should be performed in the crossed ($t$) channel, and the result crossed back to the direct ($s$) channel. 
Because of the many helicity configurations the crossing relations are complicated. 
Also, Regge exchanges have definite quantum numbers, such as parity $P$, $G$-parity $G$, and isospin $I$, which must be taken into account, 
and because of the symmetries of the nucleon-nucleon system, any non-strange mesonic Regge exchange can contribute. 
Finally, since nucleons are fermions we need to properly take into account Fermi statistics, shown in Fig. \ref{fig:stat_wFI}.

Fortunately we can can either avoid, or at least simplify, these complications by relating the Regge exchanges to the Fermi invariants \cite{gribov,sharp},
\begin{align}\label{eq:FI}
   \hat{T} &= {F_{S}^I(s,t)}1^{(1)} \cdot 1^{(2)} - {F_P^I(s,t)} (i\gamma_5)^{(1)} \cdot (i\gamma_5)^{(2)} \nonumber \\
           &+ {F_V^I(s,t)} \gamma^{\mu(1)} \gamma_{\mu}^{(2)} + {F_A^I(s,t)} (\gamma_5\gamma^{\mu})^{(1)} (\gamma_5\gamma_{\mu})^{(2)}  \\
           &+ {F_T^I(s,t)} \sigma^{\mu \nu (1)} \sigma_{\mu \nu}^{(2)}  \nonumber
\end{align}
where $s$ and $t$ are the Mandelstam variables, $I$ is an isospin label, $1$ and $2$ correspond to the vertices shown in Fig. \ref{fig:stat_wFI}.
This immediately benefits us since we get all spin dependence ``out in the open'', and we can immediately do the Dirac algebra to get the s-channel helicity amplitudes, \begin{align} \label{eq:Tpp}
T_{i}^{pp \rightarrow pp} &= \sum_j \left\{  C_{ij}^{t} \left[ F_{j}^{0}(s,t) + F_{j}^{1}(s,t) \right] 
                                           -C_{ij}^{u} \left[ F_{j}^{0}(s,u) + F_{j}^{1}(s,u) \right]  \right\} \\\label{eq:Tpn}
T_{i}^{pn \rightarrow pn} &= \sum_j \left\{  C_{ij}^{t} \left[ F_{j}^{0}(s,t) - F_{j}^{1}(s,t) \right] 
                                           -2C_{ij}^{u} F_{j}^{1}(s,u)  \right\} ,
\end{align}
where $i$ corresponds to the helicity configurations $(++;++)$, $(++;+-)$, $(+-;+-)$, $(++;--)$, $(+-;-+)$, and $j$ to the different types of Fermi invariants $ S,V,T,P,A$. The matrices $C^{t}_{ij}$ and $C^u_{ij}$, containing all the spin dependence, are obtained from performing the Dirac algebra, and are given in the appendix. For convenience Mandelstam $u$ is used in the terms corresponding to the interchange of the final state particles necessary to account for Fermi statistics, Fig. \ref{fig:stat_wFI}(b).

Our goal, instead of Reggeizeing the helicity amplitudes directly, is to instead parameterize the Fermi invariants in terms of Regge exchanges. 
This is extremely beneficial since the crossing relations become trivial as the Fermi invariants are invariant, and since we have taken care of spin explicitly,
 we will see that the Regge analysis will reduce to the spinless case.
\subsection{Relation Between Fermi Invariants and Regge Exchanges}
Regge exchanges are found in the crossed channel and have definite quantum numbers $P$, $G$, and $I$. 
$I$ is easily factored out and is taken care of in (\ref{eq:Tpp}) and (\ref{eq:Tpn}), 
so we simply need to retain the label here. Also, we need only concern ourselves with exchanges related to Fig. \ref{fig:stat_wFI}(a), 
since Reggeization of Fig. \ref{fig:stat_wFI}(b) can easily be obtained by interchanging $t \leftrightarrow u$ in our final result.
In order to find the Regge contributions to the nucleon-nucleon ($NN \rightarrow NN$) system, we need to analyze the $t$-channel, 
nucleon-anti-nucleon ($N\bar{N} \to N\bar{N}$) amplitudes of definite $P$ and $G$.

We will focus only on the initial state since there is a simple relation between the initial and final states. 
Our basic state that we will work with is,
\begin{equation} \label{eq:psi_in}
 (\psi_{in})_{\alpha \beta} = \bar v_{\alpha}(-\vec{p_t},\lambda_2) u_{\beta}(\vec{p_{t}},\lambda_1) |I,M_I  \rangle,
\end{equation}
where $p_t$ is the $t$-channel center of mass momentum, $\lambda_1$ and $\lambda_2$ are the helicities of particles 1, and 2, 
$I$ is the total isospin of the state, $M_I$ is the $3^{rd}$ component of isospin, and we have labeled the Dirac indices explicitly with $\alpha$ and $\beta$. 
The goal is to symmetrize this state in terms of parity and $G$-parity. We begin with parity. In Dirac space the parity operator is $\gamma_0$. 
Parity acting on this state yields,
\begin{equation}
  \hat{P} (\psi_{in})_{\alpha \beta} = -\bar{ v}_{\alpha}(\vec{p_t},-\lambda_2) u_{\beta}(-\vec{p_{t}},-\lambda_1) |I,M_I \rangle
\end{equation}
We can construct a vertex with definite parity then as,
\begin{equation}\label{eq:def_P_expanded}
 (\psi^{P}_{in})_{\alpha \beta}\Gamma^{(1)}_{\alpha \beta} = \frac{1}{\sqrt{2}}\left[ \bar v(-\vec{p_t},\lambda_2)\Gamma^{(1)} u(\vec{p_{t}},\lambda_1) -P \bar{ v}(\vec{p_t},-\lambda_2) \Gamma^{(1)} u_{\beta}(-\vec{p_{t}},-\lambda_1) \right] |I,M_I \rangle
\end{equation}
Defining $\gamma_0 \Gamma \gamma_0 = P_{\Gamma}\Gamma$, where $P_{\Gamma}$ is $\pm1$, we can simplify (\ref{eq:def_P_expanded}) to,
\begin{align}\label{eq:def_P}
 (\psi^{P}_{in})_{\alpha \beta}\Gamma^{(1)}_{\alpha \beta} = \frac{1}{\sqrt{2}} (1+P_{\Gamma}P) \bar v(-\vec{p_t},\lambda_2)\Gamma^{(1)} u(\vec{p_{t}},\lambda_1) |I,M_I \rangle
\end{align}
We now move on to $G$-parity. $G$-parity is defined as, $\hat{G} = \hat{\mathcal{C}}e^{i \pi \hat{I}_2} $, 
where $\hat{I}_2$ is the $y$ rotation matrix in isospin space and $\hat{\mathcal{C}}$ is the charge conjugation operator, 
given in Dirac space as, $\hat{\mathcal{C}} = \mathcal{C} \gamma_0 K$, where $K$ is the complex conjugation operator, and 
\begin{equation}
\mathcal{C} = \left( \begin{array}{cc} 0 & -i\sigma_2 \\ -i\sigma_2 & 0 \end{array} \right), \quad 
\sigma_2 = \left( \begin{array}{cc} 0 & -i \\ i & 0 \end{array} \right). 
\end{equation}
acting on a two particle state, (\ref{eq:psi_in}), yields,
\begin{equation}
 \hat{G} (\psi_{in})_{\alpha \beta} =  \bar{u}_{\alpha}(-\vec{p_t},\lambda_2) v_{\beta}(\vec{p_{t}},\lambda_1) (-1)^I\eta_{\lambda_2}\eta_{\lambda_1} \eta_C |I,M_I \rangle
\end{equation}
where $\eta_{\pm \lambda} = (-1)^{1/2 - \lambda}$, and $\eta_C$ is an arbitrary phase which is convenient to define as $\eta_C = (-1)^I\eta_{\lambda_2}\eta_{\lambda_1}$. 
We can then construct a vertex of definite parity and $G$-parity as,
 
\begin{align}
 (\psi^{PG}_{in})_{\alpha \beta}\Gamma^{(1)}_{\alpha \beta} =  \frac{1}{{2}} (1+P_{\Gamma}P)[
             \bar v(-\vec{p_t},\lambda_2)\Gamma^{(1)}u(\vec{p_{t}},\lambda_1) + G  \bar{u}(-\vec{p_t},\lambda_2)\Gamma^{(1)} v(\vec{p_{t}},\lambda_1)]|I,M_I \rangle
\end{align}
Defining $C_\Gamma \Gamma =  \mathcal{C} \gamma_0 K \Gamma  \mathcal{C} \gamma_0 K $, where $C_{\Gamma}$ is $\pm1$, this simplifies to,
\begin{equation}\label{eq:def_PG}
 (\psi^{PG}_{in})_{\alpha \beta}\Gamma^{(1)}_{\alpha \beta} = \frac{1}{2}(1+P_{\Gamma}P)(1+\eta_{\lambda_1}\eta_{\lambda_2}\mathcal{C}_{\Gamma}G)\bar v(-\vec{p_t},\lambda_2)\Gamma^{(1)} u(\vec{p_{t}},\lambda_1) |I,M_I \rangle.
\end{equation}
$P_{\Gamma}$ and $\mathcal{C}_{\Gamma}$ for the available couplings are given in Table \ref{ta:gamma_sym}. Note that we use the decomposition $\sigma^{\mu \nu (1)} \sigma^{(2)}_{\mu \nu} = -2 \vec{\alpha}^{(1)} \cdot \vec{\alpha}^{(2)} +2 \vec{\Sigma}^{(1)} \cdot \vec{\Sigma}^{(2)}$. 

\begin{table}
\caption{Symmetries of $\gamma$ matrices.}
 \begin{tabular}{ c c c c c c c c c}
 \hline \hline
$\Gamma$            &$I$ & $\gamma^{5}$ & $\gamma^{0}$ & $\vec{\gamma}$ & $\gamma^{0}\gamma^{5} $ & $\vec{\gamma}\gamma^{5}$& $i\vec{\alpha}$& $\vec{\Sigma}$  \\ 
\hline
$P_\Gamma$          & +  &     -        &     +        &      -         &           -             &            +            &       -        &        +         \\
$\mathcal{C}_\Gamma$& +  &     -        &     -        &      -         &           +             &            +            &       -        &	      -         \\
	
\hline  \hline
\end{tabular}
\label{ta:gamma_sym}
\end{table}
The result, (\ref{eq:def_PG}) can also be used to calculate outgoing states of definite parity and $G$-parity by utilizing the relations, 
\begin{align}
\left(\bar{u}(\vec{p},\lambda)\Gamma v(-\vec{p},\lambda_2) \right)^{\ast} = v^{\dag}(-\vec{p},\lambda_2)  \Gamma^{\dag} \bar{u}^{\dag}(\vec{p},\lambda_1) = \bar{v}(-\vec{p},\lambda_2)\Gamma u(\vec{p},\lambda_1).
\end{align}

With these results we can construct the symmetrized amplitudes,  
\begin{equation} \label{eq:FI_Dirac_coeef}
 \tilde{T}^{PGI}_{\lambda_1',\lambda_2';\lambda_1,\lambda_2} = S^{PG}_{\lambda_{i}} F_{S}^I(s,t) + P^{PG}_{\lambda_{i}} F_P^I(s,t) + V^{PG}_{\lambda_{i}} F_V^I(s,t)  + A^{PG}_{\lambda_{i}} F_A^I(s,t) + T^{PG}_{\lambda_{i}} F_T^I(s,t),
\end{equation}
where $\lambda_i$ represents $\lambda_1$, $\lambda_2$, $\lambda_1'$, $\lambda_2'$, 
which are the helicities of the initial and final particles, and 
$S^{PG}_{\lambda_{i}},P^{PG}_{\lambda_{i}},V^{PG}_{\lambda_{i}},A^{PG}_{\lambda_{i}},T^{PG}_{\lambda_{i}}$ are obtained from the dirac algebra, 
and are dependent upon specific values of $P$, $G$, and helicity configurations. %The details of this procedure are given in the appendix.

Now, in order to Reggeize, we set up a partial wave expansion, of definite parity and $G$-parity, in the $t$-channel, center of momentum frame,
\begin{align} \label{eq:PWE}
    \tilde{T}^{PGI}_{\lambda_1',\lambda_2';\lambda_1,\lambda_2} = 
   \sum_J(2J+1) [f^{GIJ}_{\lambda_{1}', \lambda_{2}' ; \lambda_1, \lambda_2}(E_t)
  - P(-1)^{J+\lambda}f^{GIJ}_{\lambda_{1}', \lambda_{2}' ; -\lambda_1, -\lambda_2}(E_t) ]d^J_{\lambda_1 - \lambda_1',\lambda_2 - \lambda_2'}(\theta_t),
\end{align}
where $E_t$ and $\theta_t$ are the $t$-channel center of momentum energy and scattering angle, 
and the $f^{GIJ}_{\lambda_{1}', \lambda_{2}' ; \lambda_1, \lambda_2}(E_t)$ 
correspond to partial wave coefficients from expanding on to the Wigner-d functions. 

We now equate (\ref{eq:FI_Dirac_coeef}) and (\ref{eq:PWE}) and select specific $P$, $G$ and helicity values to isolate and solve for each of the Fermi invariants 
in terms of partial waves. Essentially, there are many redundant equations since the Fermi invariants do not depend on spin, 
giving us the freedom to only choose helicity combinations of $(++;++)$ and $(+-;+-)$. Once this is done all the Wigner-d functions reduce to Legendre polynomials, 
and the Reggeization procedure reduces to the spinless case. Each Fermi invariant trivially crosses back to the s-channel, (\ref{eq:Tpp}) and (\ref{eq:Tpn}), 
and we do not have to deal with any of the complications that are associated with a typical Regge analysis of particles with spin. 
From here we can follow the typical methods to Reggeize spinless amplitudes\cite{martin,perl,collins}, the result is,
\begin{equation} \label{eq:FI_to_RP_t}
\left( \begin{array}{c}
F^{I}_{S}(s,t)  \\ \\
F^{I}_{V}(s,t) \\ \\
F^{I}_{T}(s,t) \\ \\
F^{I}_{P}(s,t) \\ \\
F^{I}_{A}(s,t)      
\end{array} \right)
=
\left( \begin{array}{ccccc}
\frac{m^2}{2(t-4m^2)} & 0 & 0 & 0 & 0 \\ \\
0 & \frac{t-4m^2}{8\left(2s+t-4m^2 \right)} & \frac{t}{8(2s+t-4m^2)} & 0 & 0 \\ \\
0 & 0 & -\frac{m^2}{4(2s+t-4m^2)} & 0 & 0 \\ \\
0 & 0 & 0 & \frac{-m^2}{2t} & 0 \\ \\
0 & 0 & 0 & 0 & \frac{1}{8} \\ \\
\end{array} \right)
\left( \begin{array}{c}
R^{I++}_{+1}(s,t)  \\ \\
R^{I--}_{-2}(s,t) \\ \\
R^{I+-}_{-3}(s,t) \\ \\
R^{I--}_{+4}(s,t) \\ \\
R^{I-+}_{+5}(s,t)      
\end{array} \right)
\end{equation}
where $m = .93895$ (GeV) is the nucleon mass, and the right-most vector is defined by a sum of Regge exchanges,
\begin{equation}\label{eq:ReggeExchange}
 R^{IPG}_{\pm j}(s,t)= \zeta(s,t)\sum_{k}\xi_{k\pm}(t)\beta_{k}^{IPG}(t) \left(-1+\frac{2s}{4m^2-t}\right)^{\alpha_k(t)},
\end{equation}  
where $\beta(t)$ and $\alpha(t)$ correspond to the residue and the trajectory of the Regge pole and are discussed in the following section, 
$\zeta(s,t)$ is a cutoff factor also discussed in the following section,
$j$ is simply the position of the Regge exchange in the vector, and 
\begin{equation}\label{eq:signature} %\frac{e^{-i\pi\alpha(t)} \pm 1}{\sin{\pi \alpha(t)}} \approx 
\xi_{\pm}(t) = \left\{ \begin{array}{c}
 e^{-i(\pi\alpha(t)/2 + \delta)} \quad  + \\ \\ -ie^{-i(\pi\alpha(t)/2 + \delta)} \quad  - 
\end{array}  \right. 
\end{equation}

Here we have also introduced an additional phase for each Regge exchange, $\delta$, which accounts for the fact that we have absorbed all extra $t$ dependence 
into the residue, including the approximation we utilized for $\xi_{\pm}(t)$. Ultimately it provides an extra degree of freedom which is convenient when fitting 
certain Reggeons. Also note that while Reggeons with $PG = --$, enter into two different positions in (\ref{eq:FI_to_RP_t}), 
the residues of any contributing poles in these positions are not necessarily the same. 

In order to take into account the $u$-channel exchanges of Fig. \ref{fig:stat_wFI}(b), we can simply substitute $t \rightarrow u$ in (\ref{eq:FI_to_RP_t}). 
We also utilize an additional factor of $\frac{t}{4m^2}$ for type 4 exchanges, guaranteeing that amplitude $d = 0$ at $t = 0$, 
which is required by conservation of angular momentum. 
In addition, we multiply type 5 exchanges with a factor of $\frac{4m^2}{s}$, which we assume we can factor from $F_{A}(s,t)$. 
This is necessary in order to cancel with an additional factor of $s$ in the matrix $C^{t}_{ij}$ and prevents amplitude $e$ from blowing up at large $s$. 
This seems to be a general problem with expressing the amplitudes in terms of the Fermi invariants at large $s$, and $F_{A}(s,t)$ should either always be 
redefined or parameterized to explicitly cancel this factor of $s$ in order to avoid this problem.  
Now that we have the Fermi invariants parametrized in terms of Regge exchanges 
we can plug this result into (\ref{eq:Tpp}) and (\ref{eq:Tpn}) for a Regge approximation of the $s-$channel helicity amplitudes.

\subsection{Residue and Trajectory}
We utilize linear Regge trajectories, $\alpha(t) = \alpha_0+\alpha_1t$. 
These are obtained from the meson masses available from the Particle Data Group\cite{PDG}. 
In addition to the mesonic trajectories, we also utilize ``effective'' trajectories as discussed below.

We use three different parametrizations for the residues,
\begin{align} \label{EQ:residues}
\beta_{I}(t)   &= \beta_0e^{\beta_1t} \nonumber \\
\beta_{II}(t)  &= \left(1 - e^{\gamma t}\right)\beta_0e^{\beta_1t}  \\  
\beta_{III}(t) &= \frac{t}{4m^2}\beta_0e^{\beta_1t} \nonumber
\end{align}
where $\beta_0$, $\beta_1$ and $\gamma$ are fit parameters.
We utilize the different types of residues for different Regge exchanges, as well as different fit solutions, 
based on trial and error.

Equation (\ref{eq:ReggeExchange}) differs from the usual expression in that we have kept the full expression for $\cos(\theta_t)$. 
Generally the Regge limit assumes that $\cos(\theta_t)>>1$ which impies that $s>> 4m^2 - t$. In extrapolating from the region where the SAID partial wave analysis
has been performed to higher $s$, we are violating this condition in two respects. First, data where $s$ is of the same order of magnitude as $4m^2$ are included. 
Second, in the same region there are significant data for $4m^2 - s < t <0$. So at backward angles $t$ is of the same order of magnitude as $s$. 
For this reason we keep the exact expresssion for $\cos(\theta_t)$.

We have chosen to fit the low s data using the same form as the Regge parameterization,
but in doing so it is necessary to introduce additional ``effective'' trajectories that are not derived from the meson spectrum.

A practical problem associated with fitting these forms at low $s$ is that the $u$ channel contributions necessarily overlap those from the $t$ channel.
Fitting to data near $\theta = 0^\circ$, where $t=0$ and $u = 4m^2 - s$, and near $\theta = 180^\circ$, where $u = 0$ and $t = 4m^2 - s$,
can be affected substantially by the tail of the crossed channel. This can cause the fitting procedure to become very sensitive, if not unstable.
As a result we have found it useful to introduce a cutoff factor,
\begin{equation}
	\zeta(s,t) = \left(1 - e^{20\left(\frac{t}{4m^2 - s} - 1\right)}\right),
\end{equation}
to decouple the $t$ and $u$ channel contributions at the endpoints in order to simplify the fitting procedure. 
This has no effect at large $s$ where
the two channels have no significant overlap, but is extremely useful for smaller values of $s$.
\subsection{Electromagnetic Effects}
In order to properly describe the proton-proton interaction we need to account for electromagnetic effects. We use the full proton vertex (\ref{eq:EM_vertex}),
with a one photon exchange. The one photon exchange amplitudes are given in the appendix (\ref{eq:EMamp}). 
In order to account for higher order effects we utilize a helicity-dependent constant and phase. 
Since the electromagnetic contribution is dominated by ``no flip'' and ``single flip'' contributions we redefine the one photon exchange amplitudes as follows,
\begin{align} \label{eq:EM_phase_amp}
 a'_{EM}(s,t)  = \beta_ae^{i\delta_a}a_{EM}(s,t) \\
 b'_{EM}(s,t)  = \beta_be^{i\delta_b}b_{EM}(s,t) \\
 c'_{EM}(s,t)  = \beta_ce^{i\delta_c}c_{EM}(s,t),
\end{align}
where $\beta_a$, $\beta_b$, $\beta_c$, $\delta_a$, $\delta_b$, and $\delta_c$  are fit to available polarization and differential cross section data. 
Utilizing this approach allows us to keep the electromagnetic effects under control, and smoothly fit between the Coulomb and hadronic regions. 

\section{Results}
\label{sec:results}
We present here two solutions to our fits, an unpolarized version as well as a polarized version. 
While one of our primary goals is to include all spin dependence, we also foresee applications for which an unpolarized solution will prove useful. 

In comparison with the SAID analysis, there is overlap between the two models at lower energies, 
and we expect the SAID solution to be more precise, i.e. lower $\chi^2$.
The emphasis on our fit is the ability to extrapolate to higher energies. 
Using a Regge model over the entire angular region may give less precise results than the SAID parameterization, but it does allow this extrapolation.
A further test of our results will be when we implement them into the electrodisintegration of the deuteron process, as it will allow us to see 
how much the results vary in comparison with the use of the SAID amplitudes through the kinematical region of overlap.

The data set was assembled from the SAID analysis \cite{SAIDdata}, the Durham database \cite{Durham}, 
the Cudell dataset \cite{Cudell:2005sg}, and the Particle Data Group \cite{PDG}. 
We also reference here the original papers 
\cite{BARASHENKOV,Schwaller:1979eu,MARSHALL,SUTTON,SCHWALLER,GUZHAVIN,MESHCHERYAKO,DZHELEPOV,DZHELEPOV2,ELIOFF,SMITH,IGO,Shimizu:1982dx,Jaros:1977it,Chen:1956zz,Hart:1962zz, COLETTI,COLETTI2,Longo:1962zz,Blue:1962zz,PARKS,Diddens:1962zz,TAYLOR,Ginestet:1969zx,Jenni:1977kv, BELLETTINI1, BELLETTINI2,Almeida:1969bv,CZAPEK,BLOBEL,BREITENLOHNE,Ashmore:1960zz,Jabiol:1977ku,Apokin:1976zu,Bushnin:1973yv,Ammosov:1972cx,Brick:1982dy,Brenner:1981kf,Bartenev:1972nf,Barish:1974ga,Kafka:1978pp,Firestone:1974pd,Dao:1972jb,Bromberg:1973fi,Amaldi:1972uw,Amos:1981dr,Amos:1985wx,Amaldi:1976yf,Baksay:1978sg,Carboni:1984sg,Amaldi:1978vc,Eggert:1975bd,Ambrosio:1982gq,Honda:1992kv,Baltrusaitis:1984ka,Shapiro:1965zza,CARVALHO,LAW,BUGG,Bugg:1966zz,Badier:1972ui,Galbraith:1965jk,Carroll:1974yv,Denisov:1971jb,Carroll:1975xf,Carroll:1978vq,KRUCHININ,DALKHAZAV,AZIMOV,Albers:2004iw,Shimizu1982445,Garcon:1986ni,Albrow:1970hn,Williams:1972gk,Kammerud:1971ac,Jenkins:1979nc,Fujii:1962zz,Eisner:1965zz,Ankenbrandt:1968zz,Rust:1970wp,Ambats:1974is,Clyde:1966zz,Preston:1960zz,Cork:1957zz,Alexander:1967zz,Rubinstein:1984kf,Akerlof:1976gk,Nagy:1978iw,Amaldi:1979kd,Breedon:1988kd,Faissler:1980fk,Breakstone:1984te,Bizard:1974hc,Terrien:1987jt,Perl:1969pg,Palevsky:1962zz,Miller:1971yc,Friedes:1965zz,Kreisler:1966kf,Altmeier:2004qz,Kobayashi:1991am,Bystricky:1986nj,Perrot:1987pv,Garcon:1986qp,DallaTorreColautti:1989nk,Andreev:2004bs,Cozzika:1967fz,Neal:1967jv,Marshak:1978wh,Ball:1987bh,Bell:1980rp,Diebold:1975yu,Miller:1977pm,Makdisi:1980gf,Zhurkin:1978rr,Ball:1999yy,Grannis:1966zz,Lin:1978xu,Parry:1973fj,BAREYRE,Allgower:1999ad,Allgower:1998ma,Allgower:1999ac,Ball:1999cn,Arvieux:1997vg,Deregel:1976jx,Abshire:1974ed,Klem:1977xq,Rust:1975kv,Aschman:1977rf,Borghini:1971mq,Kramer:1977pf,Corcoran:1980ew,Crabb:1977mg,Gaidot:1976kj,Okada:2006dd,Snyder:1978gs,Fidecaro:1980ee,Fidecaro:1978rb,Fidecaro:1981dk,Akchurin:1993xd,Ball:1992qn,Sakuda:1982eg,deLesquen:1999yz,Robrish:1970jw,Abolins:1973dy,Crabb:1979nh,Crabb:1982eh,Bauer:2004aw,Allgower:2000pv,Allgower:2001qs,Perrot:1988tw,Ball:1994uz,Bystricky:1985bz,Lehar:1988tx,Auer:1978zp,Auer:1983vm,Fontaine:1989ak,Auer:1982cv,Ball:1988pc,Abshire:1975bp,Ball:1994bp,Lac:1989fx,Lac:1989aj,Allgower:1998wf,Allgower:1998dy,Binz}. 
The dataset that we have collected can currently be obtained by contacting the authors.

All observables were fit simultaneously. The $\chi^2$ values are given in Table \ref{ta:chi^2}, 
and the parameter values are given in Tables \ref{ta:params_pol} and \ref{ta:params_unpol} 
for the polarized and unpolarized solutions respectively. In order to avoid the largest data set dominating the fit, 
we implemented weights which were varied in order to keep all observables on the same footing. 
This was especially useful since the proton-neutron data is so limited in comparison to the proton-proton data.
Because our data set includes differential cross section data from many sources, there is a potential problem with normalization. 
In order to correct for this we fit to the shape of the differential cross section data and allow the overall magnitude of the data to float by plus or minus 15\%. 

\begin{table}[htbp]
\caption{$\chi^2$ values for both the unpolarized and polarized solutions. The unpolarized solution is based on 6111 data points and has 132 parameters. 
The polarized solution is based on 13869 data points and has 138 parameters.}
\begin{tabular}{llccccc}
\hline \hline
 &  &  & \multicolumn{2}{c}{Unpolarized}   & \multicolumn{2}{c}{Polarized}      \\ 
Observable &  & $N$ & $\chi^2$ & $\chi^2/N$ & $\chi^2$ & $\chi^2/N$ \\ \hline
\multirow{2}{*}{$\sigma$}             & $pp$ & 181 & 151.3 & 0.8 & 160.4 & 0.9 \\ 
                                      & $pn$ & 69 & 9.3 & 0.1 & 11.1 & 0.2 \\ \hline
$\frac{d\sigma}{dt}(s>20)$            & $pp$ & 1635 & 2872.5 & 1.8 & 2853.5 & 1.7 \\ \hline
\multirow{2}{*}{$\frac{d\sigma}{dt}$} & $pp$ & 3481 & 6513.5 & 1.9 & 8353.3 & 2.4 \\ 
				      & $pn$ & 745 & 1338.4 & 1.8 & 1963.4 & 2.6 \\   \hline
\multirow{2}{*}{$P$ $(A_N)$}          & $pp$ & 3410 &  &  & 8411.2 & 2.5 \\ 
                                      & $pn$ & 508 &  &  & 1600.2 & 3.1 \\              \hline
\multirow{2}{*}{$A_{YY}$}             & $pp$ & 1587 &  &  & 7371.3 & 4.6 \\ 
                                      & $pn$ & 117 &  &  & 306.0 & 2.6 \\             \hline
\multirow{2}{*}{$A_{ZX}$}             & $pp$ & 568 &  &  & 3159.2 & 5.6 \\ 
                                      & $pn$ & 81 &  &  & 96.6 & 1.2 \\              \hline
\multirow{2}{*}{$A_{ZZ}$}             & $pp$ & 608 &  &  & 3505.9 & 5.8 \\ 
                                      & $pn$ & 89 &  &  & 229.8 & 2.6 \\          \hline
$A_{XX}$                              & $pp$ & 276 &  &  & 2616.7 & 9.5 \\        \hline
\multirow{2}{*}{$D$}                  & $pp$ & 188 &  &  & 919.5 & 4.9 \\ 
                                      & $pn$ & 37 &  &  & 111.3 & 3.0 \\           \hline
\multirow{2}{*}{$D_{T}$}              & $pp$ & 281 &  &  & 1885.5 & 6.7 \\ 
			              & $pn$ & 8 &  &  & 3.1 & 0.4 \\             \hline
total &  &  & 10885.1 & 1.8 & 43558.0 & 3.1 \\  %\slashbox{6111}
\hline \hline
\end{tabular}
\label{ta:chi^2}
\end{table}

\begin{figure}
    \includegraphics[width=17.8cm]{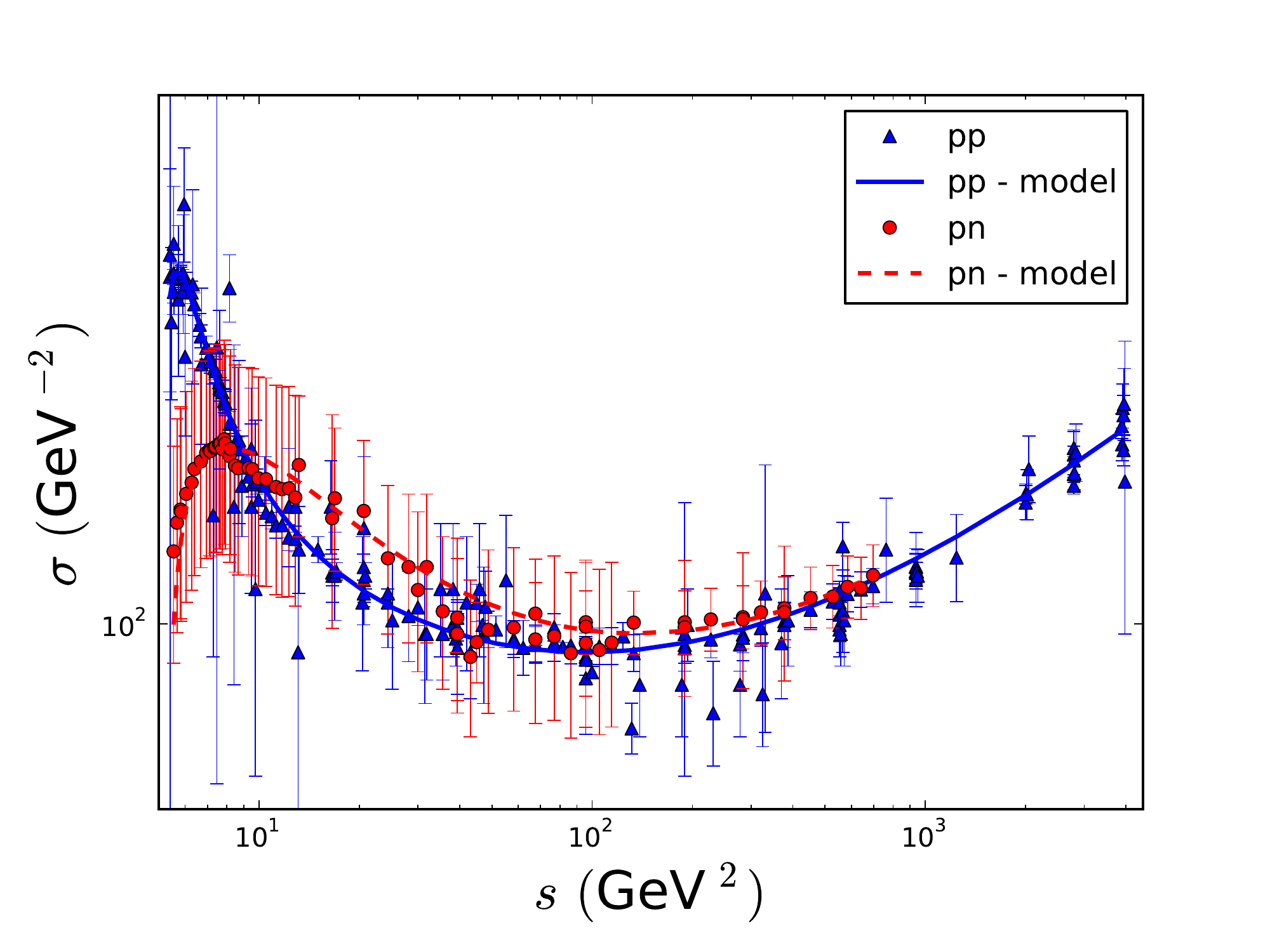}  %\textwidth
    \caption{Total cross sections for proton-proton and proton-neutron as a function of Mandelstam $s$.}
    \label{fig:total}
\end{figure}

The total cross sections for both proton-proton and proton-neutron are presented in Fig. \ref{fig:total}. 
As these are calculated at $t=0$, the Regge approximation works extremely well. 

\begin{figure}
    \includegraphics[width=17.8cm]{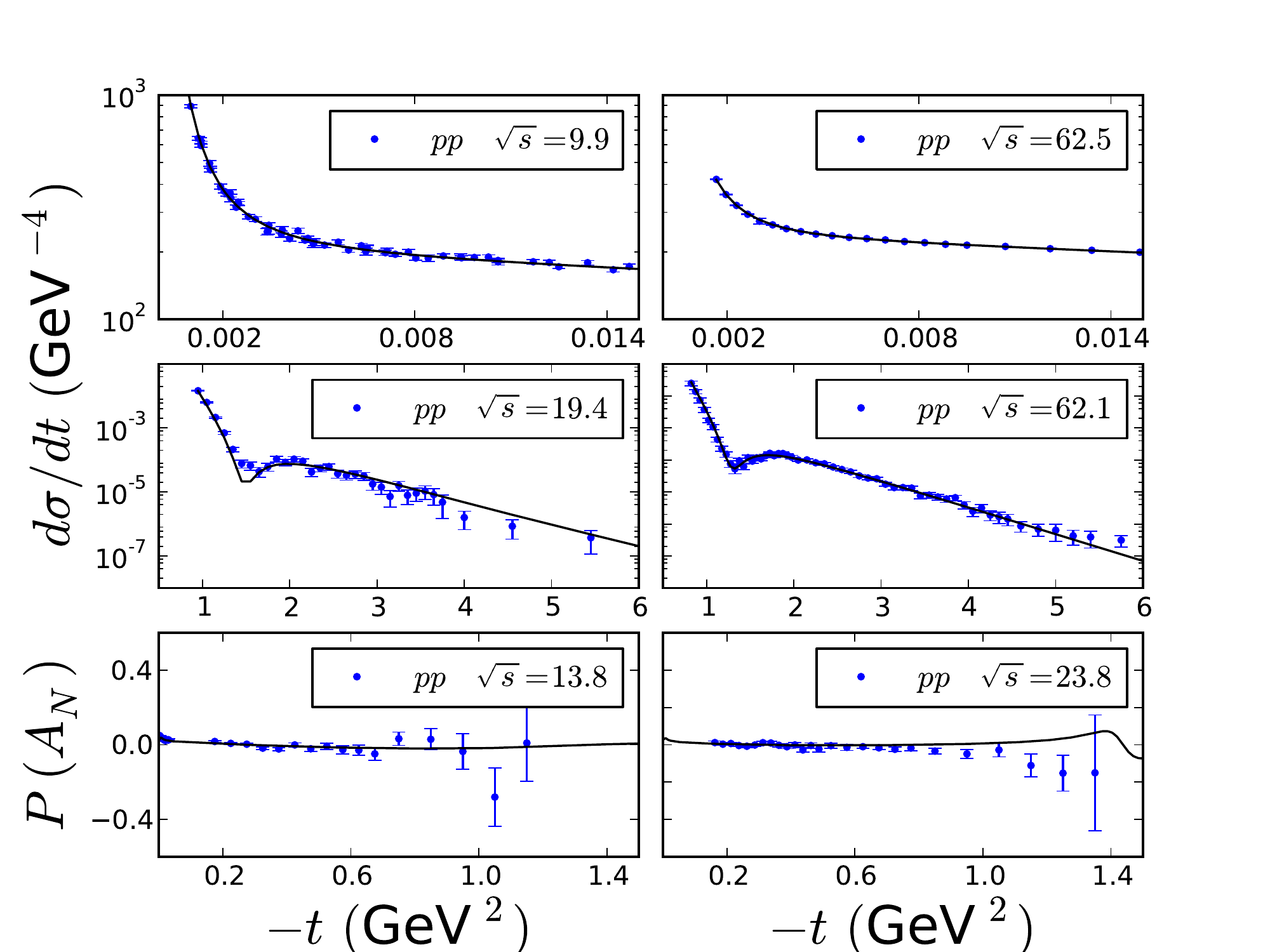} 
    \caption{High energy results for proton-proton differential cross sections in both the Coulomb and dip region, as well as high energy polarization.}
    \label{fig:Highpp}
\end{figure}

In order to constrain the model at large $s$, we also fit to high energy proton-proton data. 
In Fig. \ref{fig:Highpp} we show differential cross sections through both the Coulomb and dip regions, as well as the polarization parameter. 
These results illustrate the ability of the Regge model to scale to higher energies.

\begin{figure}
    \includegraphics[width=17.8cm]{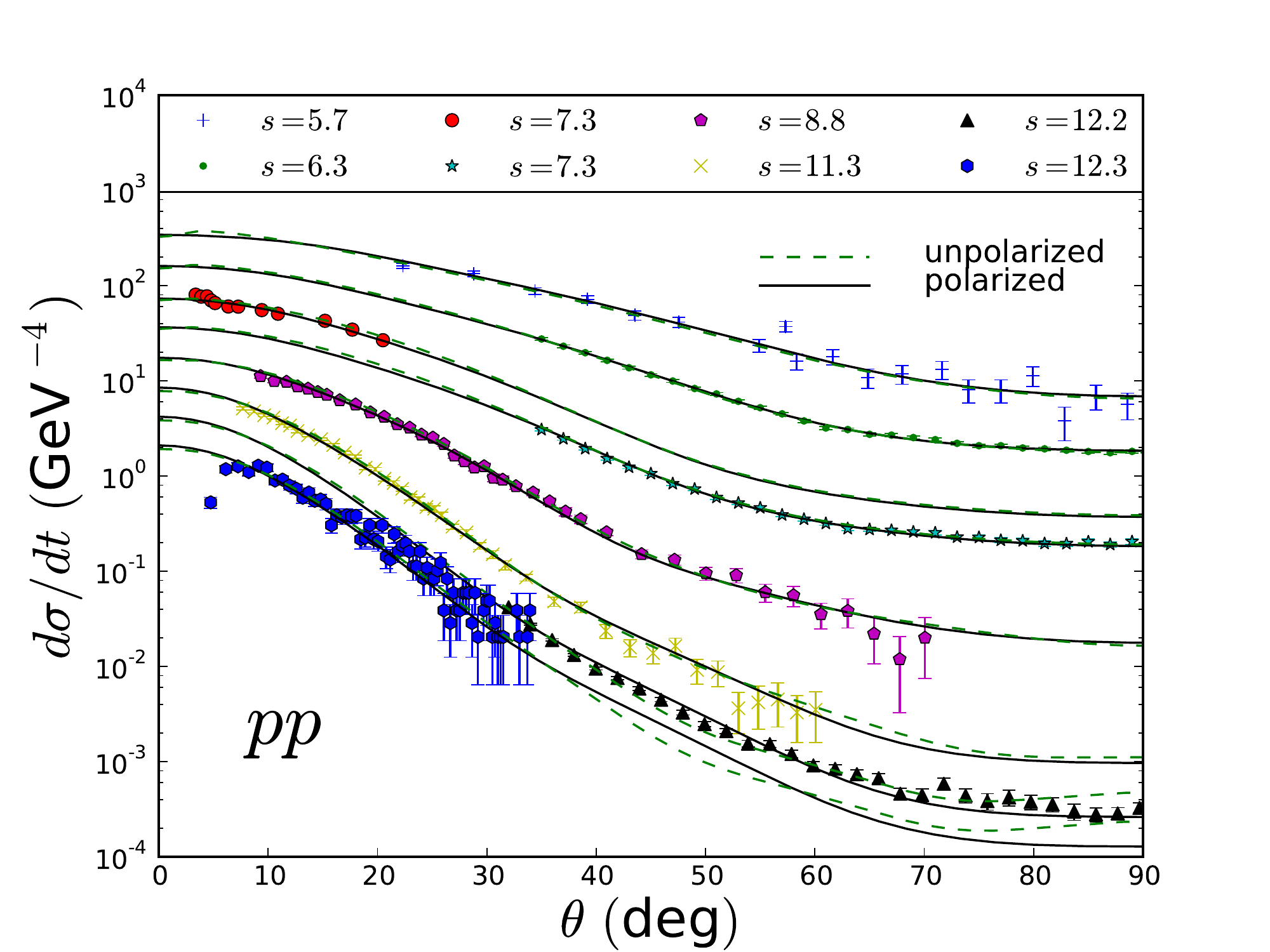} 
    \caption{Differential cross section for proton-proton as a function of center of mass angle $\theta$.
    Each data set is offset by a factor of two.}
    \label{fig:DSGppL}
\end{figure}

\begin{figure}
    \includegraphics[width=17.8cm]{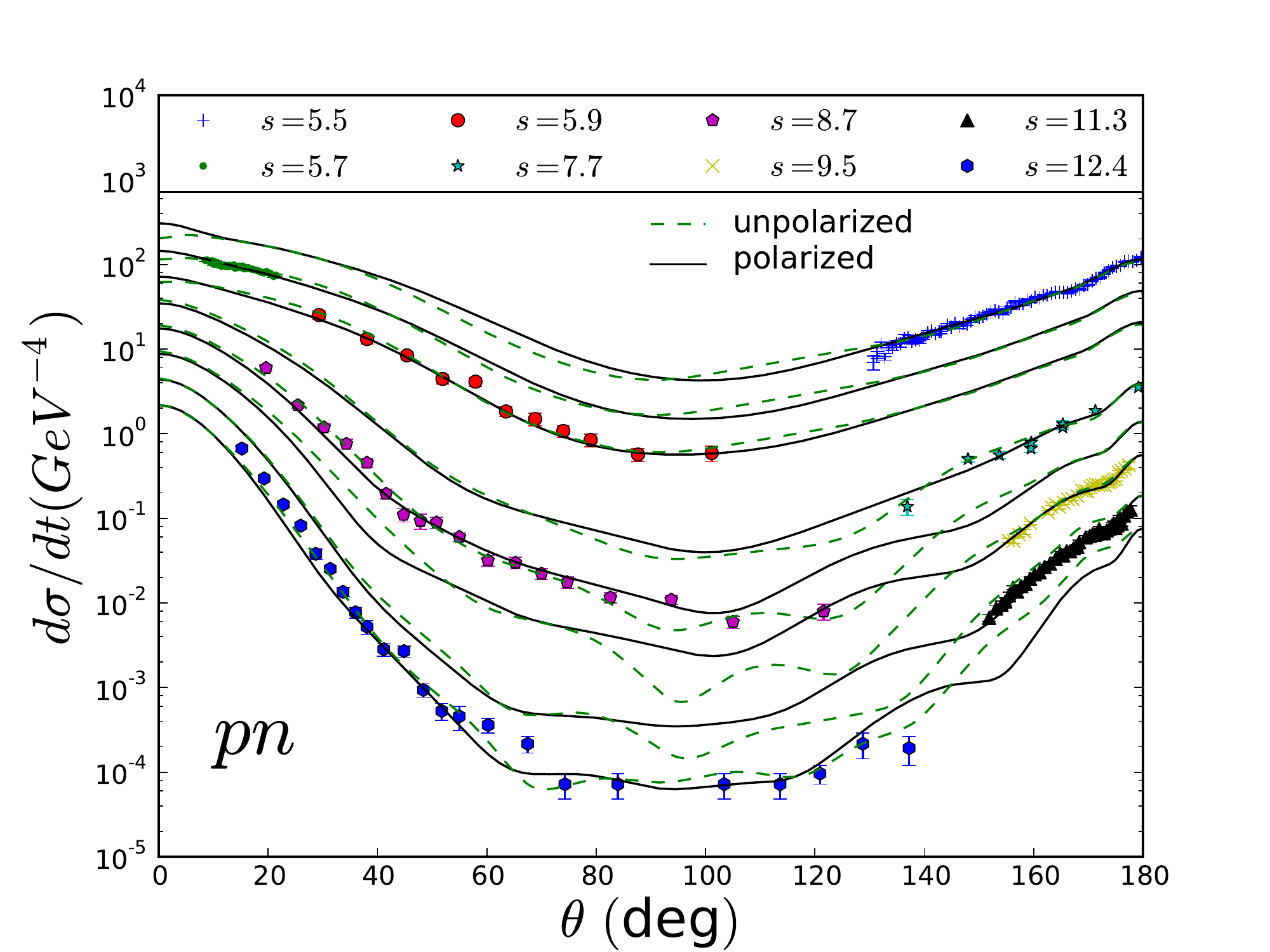} 
    \caption{Differential cross section for proton-neutron as a function of center of mass angle $\theta$. 
    Each data set is offset by a factor of two.}
    \label{fig:DSGpn}
\end{figure}

Low energy differential cross sections for proton-proton and proton-neutron are shown in Figs. \ref{fig:DSGppL} and \ref{fig:DSGpn}. 
The model works very well, especially considering that we describe the data over the entire angular region, and for relatively low $s$, 
well outside of where one would typically expect the Regge approximation to be valid. 
We have also included the unpolarized results in these graphs. While the unpolarized results yield a very low $\chi^2$, 
there is limited data to constrain the fit in the $pn$ case.

\begin{figure}
    \includegraphics[width=17.8cm]{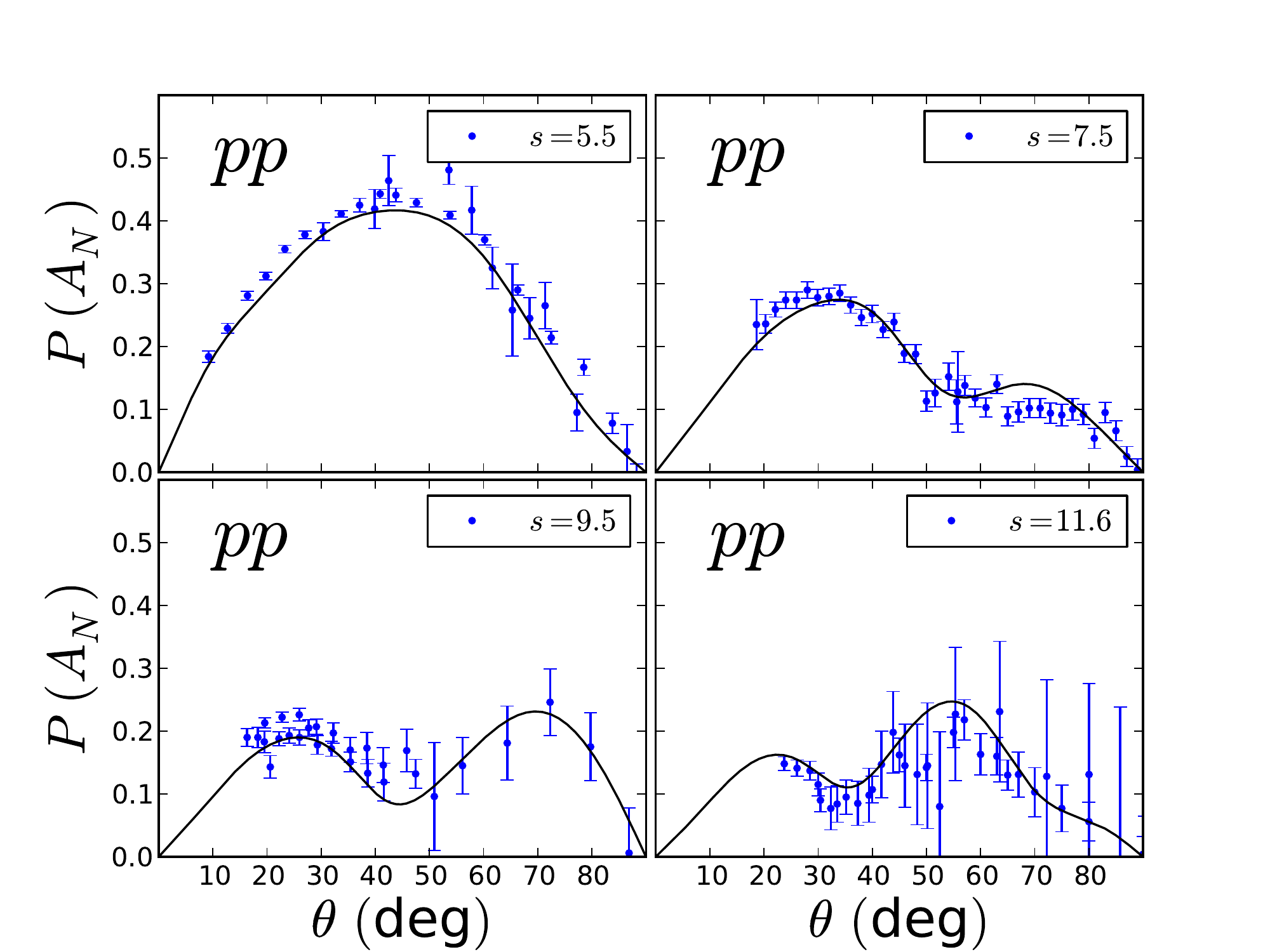} 
    \caption{Polarization for proton-proton as a function of center of mass angle $\theta$.}
    \label{fig:Ppp}
\end{figure}

\begin{figure}
    \includegraphics[width=17.8cm]{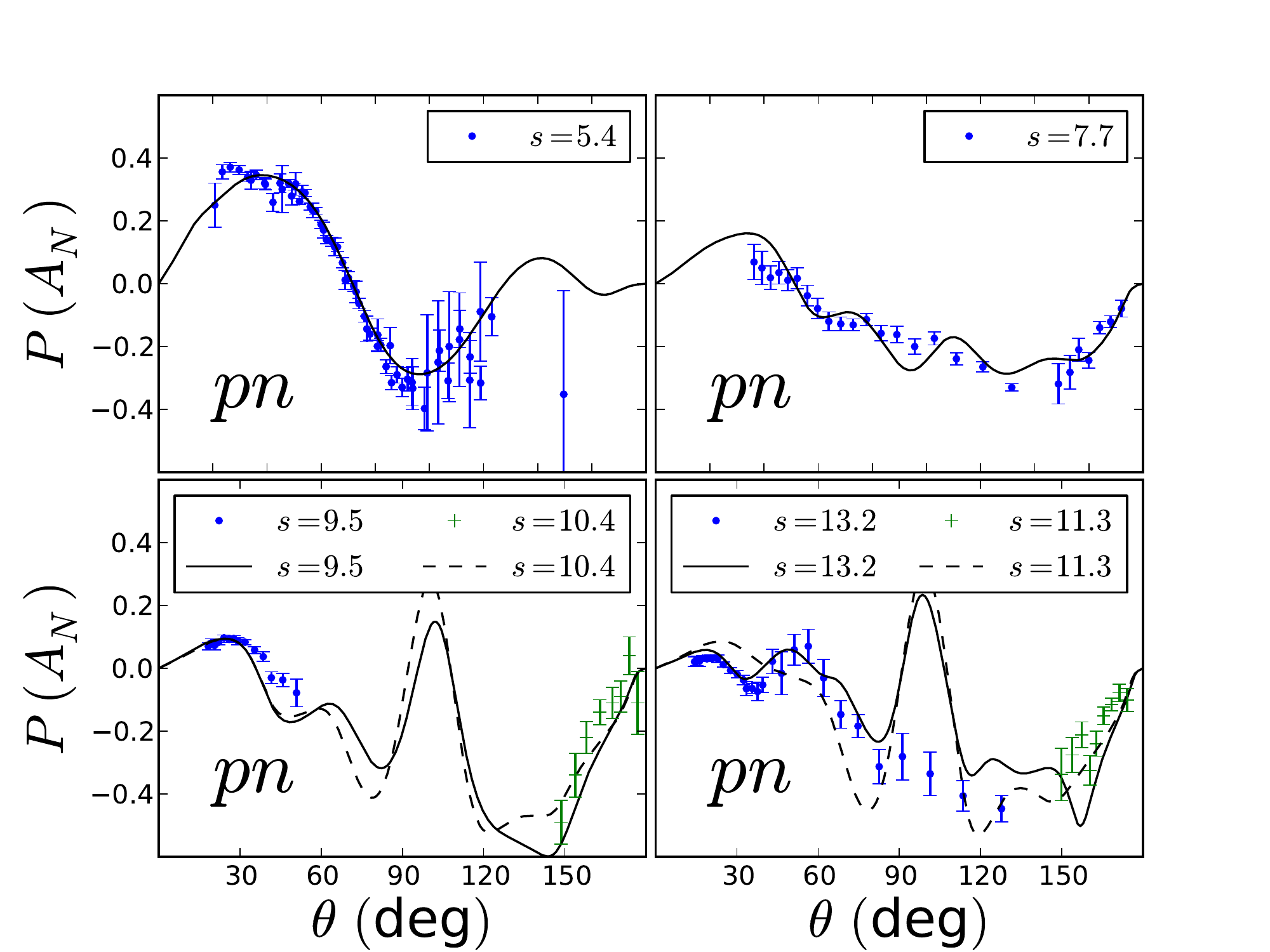} 
    \caption{Polarization for proton-neutron as a function of center of mass angle $\theta$.}
    \label{fig:Ppn}
\end{figure}

Single polarization or analyzing power are presented for proton-proton, Fig. \ref{fig:Ppp}, and for proton-neutron Fig. \ref{fig:Ppn}. 
Again the model describes the data well, although more proton-neutron data would be useful to constrain the model further. 

\begin{figure}
    \includegraphics[width=17.8cm]{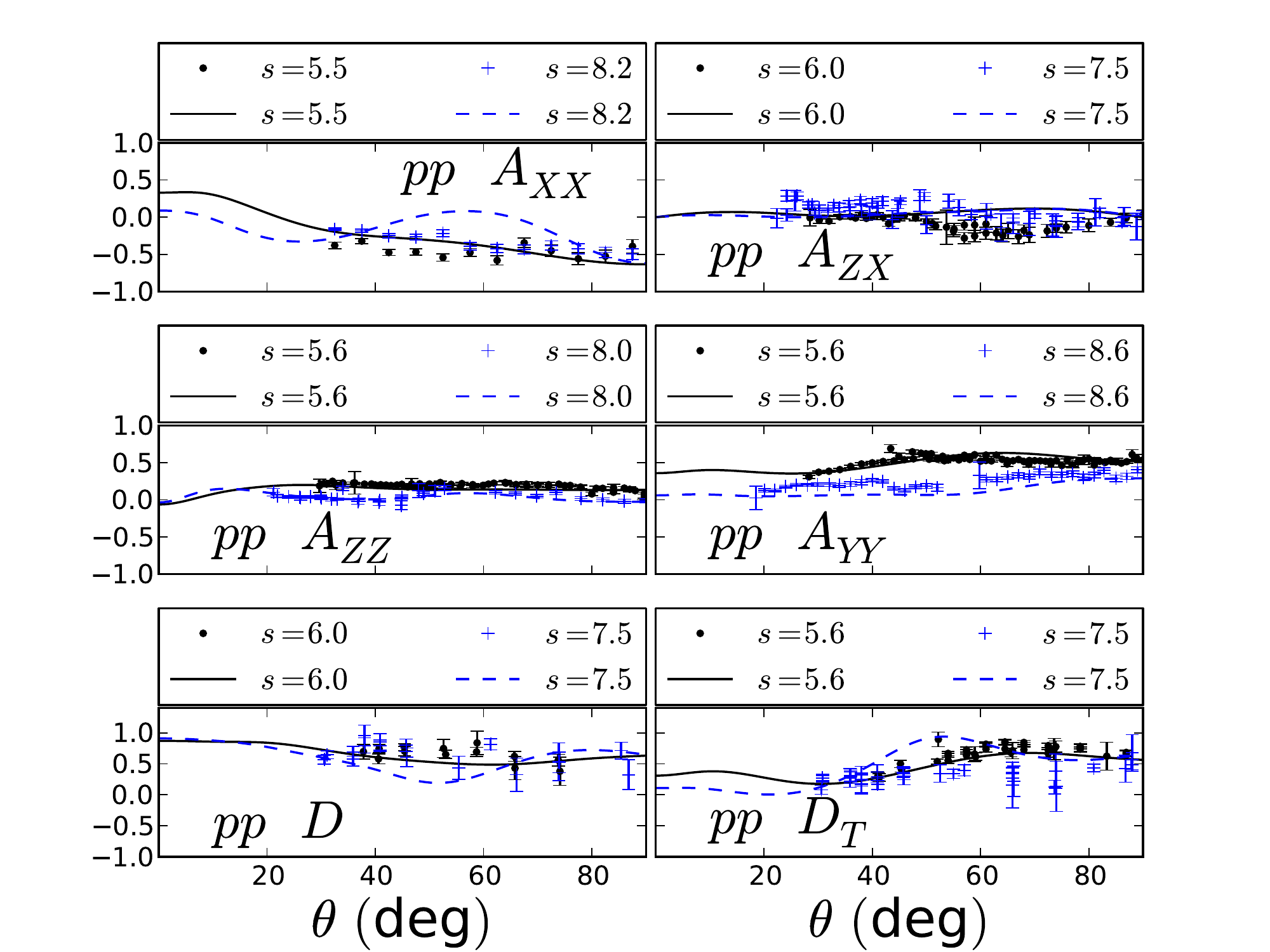} 
    \caption{Double polarization observables for proton-proton as a function of center of mass angle $\theta$.}
    \label{fig:POBSpp}
\end{figure}

\begin{figure}
    \includegraphics[width=17.8cm]{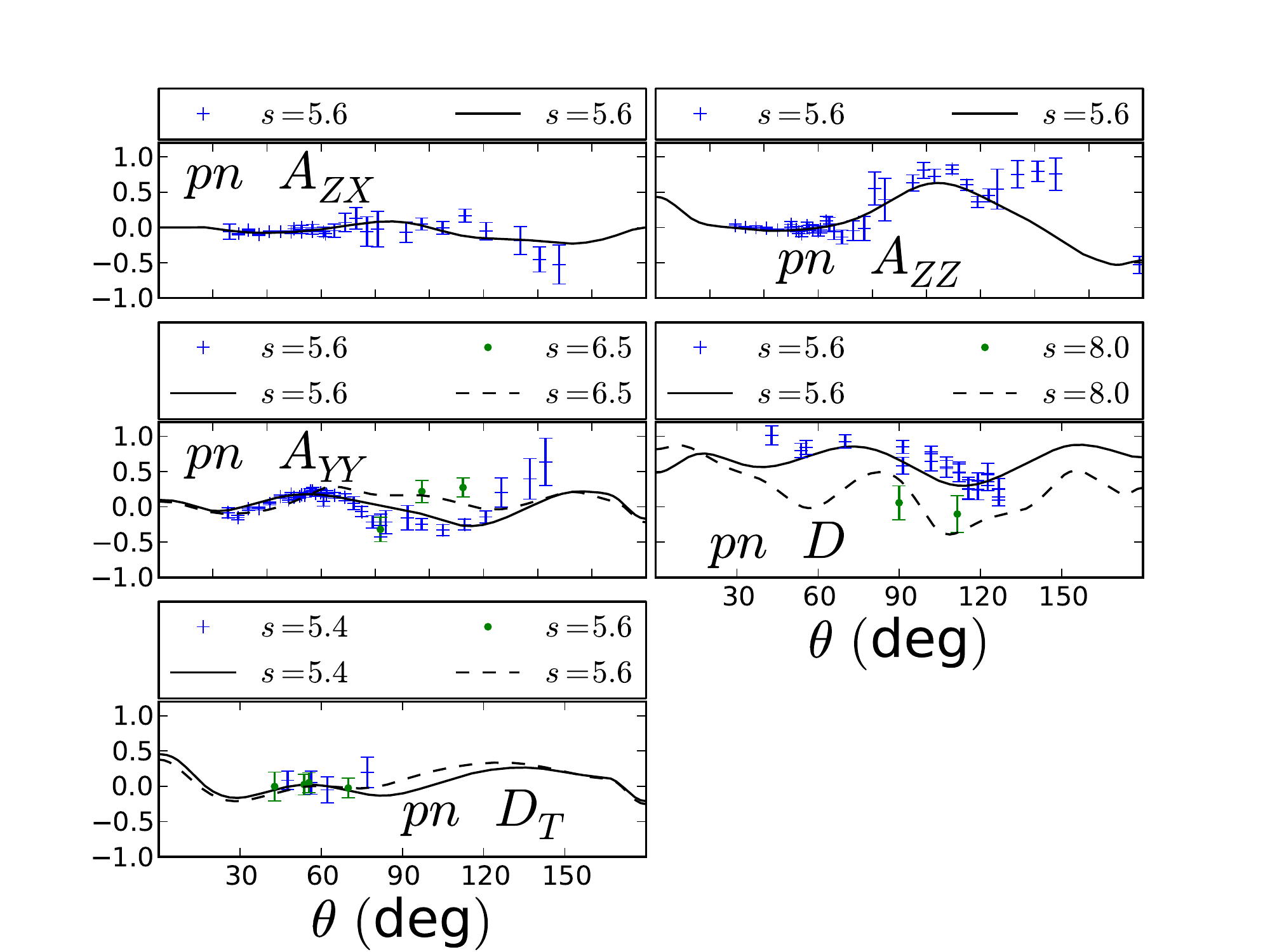} 
    \caption{Double polarization observables for proton-neutron as a function of center of mass angle $\theta$.}
    \label{fig:POBSpn}
\end{figure}

Finally we present the double-polarization observables, proton-proton in Fig. \ref{fig:POBSpp} and proton-neutron in Fig. \ref{fig:POBSpn}. 
These were fit with minimal priority, due to the lack of data. 
For each of these observables we roughly describe the data, and in certain cases the model works very well.

\section{Summary and Outlook}
\label{sec:summary}
We have parameterized the nucleon-nucleon scattering amplitudes in terms of Regge exchanges. 
Relating to the Fermi invariants allows us to calculate all spin dependence directly, 
while ensuring that a Regge exchange with definite quantum numbers contributes appropriately to the amplitude. 
Because of the ability of Regge theory to scale, we are able to extrapolate our results to $s \approx 20$ GeV$^2$ with reasonable confidence.

Our next step is to utilize these amplitudes to describe the final state interactions of the D(e,e'p)n process.
We also plan to provide these amplitudes, as well as the data set we have collected to the nuclear physics community.
Currently the amplitudes and data set can be obtained by contacting the authors.
\clearpage
\appendix
\section{Parameters}
Parameter values for polarized and unpolarized results are given in Table \ref{ta:params_pol}, and Table \ref{ta:params_unpol} respectively.
\begin{turnpage}
\begin{table}%[htbp]
\caption{Parameter values for polarized solution. The fit parameters are indicated in bold.}
\label{ta:params_pol}
%\tiny
\begin{tabular}{cccccccccccc}
\hline \hline
$\beta_0$ & $\beta_1$ & $\delta$ & $\gamma$ & $\alpha_0$ & $\alpha_1$ & Isospin & Parity & G-Parity & Type & Residue & Name  \\  
\hline
\textbf{-2.3014E+02} & \textbf{3.0982E+00} & \multicolumn{1}{r}{0.0000E+00} &  & 1.0800E+00 & 2.5000E-01 & 0 & $+$ & $+$ & 1 & $I$ & $\Pom$ \\ 
\textbf{3.3606E+01} & \textbf{2.5208E+00} & \textbf{-1.3505E+00} & 3.0000E+00 & \textbf{1.2915E+00} & \textbf{3.0031E-01} & 0 & $+$ & $+$ & 1 & $II$ & $X_1$ \\ 
\textbf{-1.4315E+00} & \textbf{4.2364E-01} & \textbf{-3.2163E+00} & \textbf{3.3330E-01} & \textbf{1.2228E+00} & \textbf{-7.6208E-02} & 0 & $+$ & $+$ & 1 & $II$ & $X_2$ \\ 
\textbf{-4.2550E+02} & \textbf{8.7558E+00} & \textbf{-2.8650E-01} &  & 6.7000E-01 & 8.2000E-01 & 0 & $+$ & $+$ & 1 & $I$ & $f$ \\ 
\textbf{3.0579E+03} & \textbf{3.5797E+00} & \textbf{-5.0940E+00} &  & \textbf{-7.1114E-01} & \textbf{-1.1570E+00} & 0 & $+$ & $+$ & 1 & $I$ & $X_3$ \\ 
\textbf{-5.1011E+01} & \textbf{-3.9362E-01} & \multicolumn{1}{r}{0.0000E+00} &  & 4.3000E-01 & 9.2000E-01 & 0 & $-$ & $-$ & 2 & $I$ & $\omega_a$ \\ 
\textbf{-8.3319E+02} & \textbf{6.0000E+00} & \textbf{-1.8189E+00} &  & 1.3000E-01 & 8.3000E-01 & 0 & $-$ & $-$ & 2 & $I$ & $\phi_a$ \\ 
\textbf{3.3968E+05} & \textbf{4.0113E+01} & \textbf{2.0543E+00} &  & \textbf{-8.3722E+00} & \textbf{-1.1658E-03} & 0 & $-$ & $-$ & 2 & $I$ & $X_4$ \\ 
\textbf{3.6954E+02} & \textbf{1.0385E+01} & \textbf{-2.8048E+00} &  & -2.3000E-01 & 8.6000E-01 & 0 & $+$ & $-$ & 3 & $I$ & $h$ \\ 
\textbf{-1.7985E+02} & \textbf{1.4258E+00} & \textbf{1.4192E+00} &  & \textbf{5.5908E-01} & \textbf{8.2000E-01} & 0 & $+$ & $-$ & 3 & $I$ & $X_5$ \\ 
\textbf{-3.2225E+01} & \textbf{-3.5689E-02} & \textbf{-4.1374E+00} &  & 4.3000E-01 & 9.2000E-01 & 0 & $-$ & $-$ & 4 & $I$ & $\omega_b$ \\ 
\textbf{-8.5937E+03} & \textbf{6.0000E+00} & \textbf{9.0817E-01} &  & 1.3000E-01 & 8.3000E-01 & 0 & $-$ & $-$ & 4 & $I$ & $\phi_b$ \\ 
\textbf{1.8226E+03} & \textbf{9.5443E-01} & \textbf{3.2746E+00} &  & \textbf{-6.5816E+00} & \textbf{8.1649E-04} & 0 & $-$ & $-$ & 4 & $I$ & $X_6$ \\ 
\textbf{-5.0967E+02} & \textbf{1.6424E+00} & \textbf{-4.0676E-01} &  & -2.3000E-01 & 8.6000E-01 & 0 & $-$ & $+$ & 5 & $I$ & $\eta$ \\ 
\textbf{3.4250E+02} & \textbf{1.2447E+00} & \textbf{-1.4211E+00} &  & \textbf{5.9469E-03} & \textbf{-2.4531E-01} & 0 & $-$ & $+$ & 5 & $I$ & $X_7$ \\ 
\textbf{7.7744E+01} & \textbf{-3.5462E+01} & \textbf{2.5834E-01} &  & -4.0000E-02 & 7.2000E-01 & 1 & $+$ & $+$ & 1 & $I$ & $b$ \\ 
\textbf{-5.9219E+02} & \textbf{1.3809E+00} & \textbf{-3.0997E+00} &  & \textbf{-5.5996E-01} & \textbf{4.4269E-01} & 1 & $+$ & $+$ & 1 & $I$ & $X_8$ \\ 
\textbf{3.4534E+02} & \textbf{-1.2989E+00} & \textbf{-5.2535E+00} &  & -4.0000E-02 & 7.2000E-01 & 1 & $-$ & $-$ & 2 & $I$ & $\pi_a$ \\ 
\hline \hline
\end{tabular}
\end{table}

\begin{table}%[htbp]
TABLE \ref{ta:params_pol}. (Continued)
%\caption{Parameter values for polarized solution (continued)}
%\tiny
\begin{tabular}{cccccccccccc}
\hline \hline
$\beta_0$ & $\beta_1$ & $\delta$ & $\gamma$ & $\alpha_0$ & $\alpha_1$ & Isospin & Parity & G-Parity & Type & Residue & Name  \\  
\hline
\textbf{9.1120E+02} & \textbf{6.8368E-01} & \textbf{1.1157E-01} &  & \textbf{-1.1372E+01} & \textbf{-1.7908E-05} & 1 & $-$ & $-$ & 2 & $I$ & $X_9$ \\ 
\textbf{6.8240E+02} & \textbf{-4.3461E-01} & \textbf{-5.3158E-01} &  & 4.7000E-01 & 8.9000E-01 & 1 & $+$ & $-$ & 3 & $I$ & $a$ \\ 
\textbf{6.6961E+02} & \textbf{5.4994E-01} & \textbf{2.3584E+00} &  & \textbf{4.5559E-01} & 8.9000E-01 & 1 & $+$ & $-$ & 3 & $I$ & $X_{10}$ \\ 
\textbf{2.9363E+02} & \textbf{4.4182E-01} & \textbf{-2.3057E+00} &  & \textbf{2.8195E-01} & \textbf{-8.9330E-01} & 1 & $+$ & $-$ & 3 & $I$ & $X_{11}$ \\ 
\textbf{-2.6347E+01} & \textbf{1.7260E-03} & \textbf{3.1945E+00} &  & -4.0000E-02 & 7.2000E-01 & 1 & $-$ & $-$ & 4 & $I$ & $\pi_b$ \\ 
\textbf{-1.5941E+03} & \textbf{1.6051E+00} & \textbf{-3.8402E+00} &  & \textbf{-1.3492E+00} & \textbf{1.5166E-04} & 1 & $-$ & $-$ & 4 & $I$ & $X_{12}$ \\ 
\textbf{1.1653E+02} & \textbf{6.0278E+01} & \textbf{-5.5234E+00} &  & \textbf{7.0679E-02} & \textbf{-8.7412E-02} & 1 & $-$ & $+$ & 5 & $I$ & $X_{13}$ \\ 
\textbf{1.3195E+04} & \textbf{4.1072E+00} & \textbf{-1.5099E+00} &  & 1.3000E-01 & 8.3000E-01 & 0 & $-$ & $-$ & 2 & $III$ & $X_{14}$  \\ 
\textbf{-1.4708E+03} & \textbf{-1.3829E+00} & \textbf{1.9703E+00} &  & -4.0000E-02 & 7.2000E-01 & 1 & $+$ & $+$ & 1 & $III$ & $X_{15}$ \\ 
\textbf{-2.2575E+03} & \textbf{-1.7071E+00} & \textbf{-2.5369E+00} &  & -4.0000E-02 & 7.2000E-01 & 1 & $-$ & $-$ & 2 & $III$ & $X_{16}$ \\ 
\textbf{4.2280E+03} & \textbf{-1.0389E+00} & \textbf{-5.7058E-01} &  & \textbf{-1.8954E+00} & \textbf{2.0775E-01} & 0 & $+$ & $+$ & 1 & $III$ & $X_{17}$ \\ 
\textbf{-6.3292E+02} & \textbf{7.2339E-01} & \textbf{5.0303E+00} &  & \textbf{-6.0089E+00} & \textbf{4.8252E-01} & 0 & $-$ & $-$ & 2 & $III$ & $X_{18}$ \\ 
\textbf{-2.5071E+04} & \textbf{1.9199E+00} & \textbf{2.3593E-01} &  & \textbf{-1.2202E+01} & \textbf{-1.0805E-05} & 1 & $+$ & $+$ & 1 & $III$ & $X_{19}$ \\ 
\textbf{9.6450E+02} & \textbf{7.5662E-01} & \textbf{1.1350E-01} &  & \textbf{-4.4005E+00} & \textbf{4.4305E-03} & 0 & $+$ & $-$ & 3 & $III$ & $X_{21}$ \\ 
\textbf{4.6080E+01} & \textbf{1.5845E-01} & \multicolumn{1}{r}{0.0000E+00} &  & \textbf{-2.1192E-01} & \textbf{3.7327E-01} & 0 & $-$ & $-$ & 4 & $III$ & $X_{23}$ \\ 
\textbf{1.3204E+00} & \textbf{} & \textbf{3.0270E+00} &  &  &  &  &  &  &  &  & $EM_a$ \\ 
\textbf{5.2979E-01} & \textbf{} & \textbf{3.1599E+00} &  &  &  &  &  &  &  &  & $EM_b$ \\ 
\textbf{4.6382E-01} & \textbf{} & \textbf{3.7668E+00} &  &  &  &  &  &  &  &  & $EM_c$ \\
\hline \hline
\end{tabular}
\end{table}
\end{turnpage}

\begin{turnpage}
\begin{table}[htbp]
\caption{Parameter values for unpolarized solution. The fit parameters are indicated in bold.}
\label{ta:params_unpol}
\begin{tabular}{cccccccccccc}
\hline \hline
$\beta_0$ & $\beta_1$ & $\delta$ & $\gamma$ & $\alpha_0$ & $\alpha_1$ & Isospin & Parity & G-Parity & Type & Residue type & Name \\ \hline 
\textbf{-2.3014E+02} & \textbf{3.0982E+00} & 0.0000E+00 &  & 1.0800E+00 & 2.5000E-01 & 0 & $+$ & $+$ & 1 & $I$ & $\Pom$ \\ 
\textbf{3.3606E+01} & \textbf{2.5208E+00} & \textbf{-1.3505E+00} & 3.0000E+00 & \textbf{1.2915E+00} & \textbf{3.0031E-01} & 0 & $+$ & $+$ & 1 & $II$ & $X_1$ \\ 
\textbf{-1.4315E+00} & \textbf{4.2364E-01} & \textbf{-3.2163E+00} & \textbf{3.3330E-01} & \textbf{1.2228E+00} & \textbf{-7.6208E-02} & 0 & $+$ & $+$ & 1 & $II$ & $X_2$ \\ 
\textbf{-4.2550E+02} & \textbf{8.7558E+00} & \textbf{-2.8650E-01} &  & 6.7000E-01 & 8.2000E-01 & 0 & $+$ & $+$ & 1 & $I$ & $f$ \\ 
\textbf{2.6491E+03} & \textbf{4.4051E+00} & \textbf{-3.4582E+00} &  & \textbf{-1.7172E+00} & \textbf{-2.9128E-05} & 0 & $+$ & $+$ & 1 & $I$ & $X_3$ \\ 
\textbf{-3.2822E+01} & \textbf{-1.4253E-01} & \textbf{-7.7596E-02} &  & 4.3000E-01 & 9.2000E-01 & 0 & $-$ & $-$ & 2 & $I$ & $\omega_a$ \\ 
\textbf{2.1614E+02} & \textbf{9.2400E+02} & \textbf{1.0919E+00} &  & 1.3000E-01 & 8.3000E-01 & 0 & $-$ & $-$ & 2 & $I$ & $\phi_a$ \\ 
\textbf{4.1651E+04} & \textbf{3.7156E+00} & \textbf{6.8897E-01} &  & \textbf{-7.8837E+00} & \textbf{-1.9233E+00} & 0 & $-$ & $-$ & 2 & $I$ & $X_4$ \\ 
\textbf{2.0270E+02} & \textbf{1.0385E+01} & \textbf{-4.2739E+00} &  & -2.3000E-01 & 8.6000E-01 & 0 & $+$ & $-$ & 3 & $I$ & $h$ \\ 
\textbf{-1.2879E+00} & \textbf{-8.3758E-06} & \textbf{8.9998E-01} &  & \textbf{1.3844E+00} & 8.2000E-01 & 0 & $+$ & $-$ & 3 & $I$ & $X_5$ \\ 
\textbf{1.9353E+01} & \textbf{-3.5689E-02} & \textbf{-5.4629E+00} &  & 4.3000E-01 & 9.2000E-01 & 0 & $-$ & $-$ & 4 & $I$ & $\omega_b$ \\ 
\textbf{-1.7345E+04} & \textbf{6.0000E+00} & \textbf{1.2766E+00} &  & 1.3000E-01 & 8.3000E-01 & 0 & $-$ & $-$ & 4 & $I$ & $\phi_b$ \\ 
\textbf{1.3702E+03} & \textbf{5.8600E-01} & \textbf{5.2754E+00} &  & \textbf{-4.4008E+00} & \textbf{-6.6653E-01} & 0 & $-$ & $-$ & 4 & $I$ & $X_6$ \\ 
\textbf{-2.9867E+01} & \textbf{-2.0935E-03} & \textbf{3.0541E-04} &  & -2.3000E-01 & 8.6000E-01 & 0 & $-$ & $+$ & 5 & $I$ & $\eta$ \\ 
\textbf{7.7053E+01} & \textbf{2.4688E-05} & \textbf{-1.1493E+00} &  & \textbf{5.4644E-01} & \textbf{-8.9621E-01} & 0 & $-$ & $+$ & 5 & $I$ & $X_7$ \\ 
\textbf{2.6850E+02} & \textbf{-3.5462E+01} & \textbf{3.1197E-01} &  & -4.0000E-02 & 7.2000E-01 & 1 & $+$ & $+$ & 1 & $I$ & $b$ \\ 
\textbf{-8.0481E+02} & \textbf{1.5700E+01} & \textbf{-2.5542E+00} &  & \textbf{-1.8343E+00} & \textbf{-4.9032E-04} & 1 & $+$ & $+$ & 1 & $I$ & $X_8$ \\ 
\hline \hline
\end{tabular}
\end{table}

\begin{table}[htbp]
TABLE \ref{ta:params_unpol}. (Continued)
\begin{tabular}{cccccccccccc}
\hline \hline
$\beta_0$ & $\beta_1$ & $\delta$ & $\gamma$ & $\alpha_0$ & $\alpha_1$ & Isospin & Parity & G-Parity & Type & Residue type & Name \\ \hline 
\textbf{2.1070E+01} & \textbf{-9.8069E-03} & \textbf{-3.1377E+00} &  & -4.0000E-02 & 7.2000E-01 & 1 & $-$ & $-$ & 2 & $I$ & $\pi_a$ \\ 
\textbf{4.0621E+01} & \textbf{5.9427E-01} & \textbf{-2.4486E-01} &  & \textbf{-6.3368E-01} & \textbf{2.9832E-04} & 1 & $-$ & $-$ & 2 & $I$ & $X_9$ \\ 
\textbf{8.3864E+02} & \textbf{-7.5961E-01} & \textbf{-3.0230E-01} &  & 4.7000E-01 & 8.9000E-01 & 1 & $+$ & $-$ & 3 & $I$ & $a$ \\ 
\textbf{5.9883E+02} & \textbf{5.4362E-01} & \textbf{2.7222E+00} &  & \textbf{5.0397E-01} & 8.9000E-01 & 1 & $+$ & $-$ & 3 & $I$ & $X_{10}$ \\ 
\textbf{4.1328E+02} & \textbf{2.5177E+00} & \textbf{-2.8391E+00} &  & \textbf{1.1444E-01} & \textbf{-1.3527E-05} & 1 & $+$ & $-$ & 3 & $I$ & $X_{11}$ \\ 
\textbf{2.2867E+01} & \textbf{1.1023E-06} & \textbf{4.6222E+00} &  & -4.0000E-02 & 7.2000E-01 & 1 & $-$ & $-$ & 4 & $I$ & $\pi_b$ \\ 
\textbf{-4.3636E+03} & \textbf{4.1122E+00} & \textbf{-3.3747E+00} &  & \textbf{-6.0418E-02} & \textbf{-2.3409E-05} & 1 & $-$ & $-$ & 4 & $I$ & $X_{12}$ \\ 
\textbf{8.0174E+01} & \textbf{5.0000E+00} & \textbf{-1.1248E-02} &  & 4.7000E-01 & 8.9000E-01 & 1 & $-$ & $+$ & 5 & $I$ & $\rho$ \\ 
\textbf{1.2615E+02} & \textbf{3.7157E+01} & \textbf{-2.3941E+00} &  & \textbf{-5.3900E-03} & \textbf{-2.5158E-01} & 1 & $-$ & $+$ & 5 & $I$ & $X_{13}$ \\ 
\textbf{1.1947E+04} & \textbf{4.1937E+00} & \textbf{-1.0997E+00} &  & 1.3000E-01 & 8.3000E-01 & 0 & $-$ & $-$ & 2 & $III$ & $X_{14}$ \\ 
\textbf{-1.5180E+01} & \textbf{1.6027E-03} & \textbf{-1.6412E-01} &  & -4.0000E-02 & 7.2000E-01 & 1 & $+$ & $+$ & 1 & $III$ & $X_{15}$ \\ 
\textbf{-1.9266E+01} & \textbf{-1.1984E-02} & \textbf{-1.2916E-01} &  & -4.0000E-02 & 7.2000E-01 & 1 & $-$ & $-$ & 2 & $III$ & $X_{16}$ \\ 
\textbf{3.1526E+02} & \textbf{-4.4662E-01} & \textbf{4.9006E-02} &  & \textbf{-1.5001E+00} & \textbf{-1.5091E-03} & 0 & $+$ & $+$ & 1 & $III$ & $X_{17}$ \\ 
\textbf{1.3989E+01} & \textbf{5.8787E-02} & \textbf{7.9986E+00} &  & \textbf{-6.1197E+00} & \textbf{4.8648E-01} & 0 & $-$ & $-$ & 2 & $III$ & $X_{18}$ \\ 
\textbf{-2.5159E+02} & \textbf{8.2800E-02} & \textbf{3.6994E+00} &  & \textbf{-8.7920E+00} & \textbf{2.5026E-04} & 1 & $+$ & $+$ & 1 & $III$ & $X_{19}$ \\ 
\textbf{1.3204E+00} & \textbf{} & \textbf{3.0270E+00} &  &  &  &  &  &  &  &  & $EM_a$ \\ 
\textbf{5.2979E-01} & \textbf{} & \textbf{3.1599E+00} &  &  &  &  &  &  &  &  & $EM_b$ \\ 
\textbf{4.6382E-01} & \textbf{} & \textbf{3.7668E+00} &  &  &  &  &  &  &  &  & $EM_c$ \\ 
\hline \hline
\end{tabular}
\end{table}

\end{turnpage}

\section{Amplitudes and Observables}
All observables can be written in terms of the five independent helicity amplitudes \cite{Bystricky} given in (\ref{eq:amplitudes(abcde)}). 
We present here the observables relevant to this paper,
\begin{align}
 \sigma &= \frac{-2m^2}{\sqrt{s(s-4m^2)}}\Im\left[ a + c \right]_{t = 0}   \\
  \frac{d\sigma}{dt} &= \frac{m^4}{2 \pi s(s-4m^2)}\left( |a|^2 +4|b|^2 +|c|^2 +|d|^2 + |e|^2 \right) \\  
 \tilde{\sigma} &=  \frac{1}{2}\left(|a|^2 +4|b|^2 +|c|^2 +|d|^2 + |e|^2 \right) \\
 \frac{d\sigma}{dt} &= \frac{m^4}{ \pi s(s-4m^2)}\tilde{\sigma}   \\ 
 \tilde{\sigma}P    &= \tilde{\sigma}A_N = -\Im[b^{*}(a + c + d - e)] \\
 \tilde{\sigma}A_{XX} &=  \Re(a^{*}d + c^{*}e) \\
 \tilde{\sigma}A_{ZX} &=  -\Re[b^{*}(a+d-c+e)] \\
 \tilde{\sigma}A_{ZZ} &= -\frac{1}{2}\left(|a|^2  +|d|^2 - |c|^2 - |e|^2 \right) \\
 \tilde{\sigma}A_{YY} &=  \Re(a^{*}d - c^*e) +2|b|^2 \\
 \tilde{\sigma}D      &=  \Re(a^{*}c - d^*e) +2|b|^2 \\
 \tilde{\sigma}D_{T}  &=  \Re(a^{*}e - d^*c) +2|b|^2
\end{align}
\section{Helicity Spinors}
In the center of momentum frame the helicity spinors are,
\begin{align} \label{eq:spinors}
 u( \pm \mathbf{p}, \lambda) &= N \left( \begin{array}{c} 1  \\  2 \lambda \tilde{p} \end{array} \right) \chi_{\pm \lambda}(\mathbf{\hat{p}}) , \\
 v( \pm \mathbf{p}, \lambda) &= N \left( \begin{array}{c} -2 \lambda \tilde{p}  \\ 1 \end{array} \right) \chi_{\mp \lambda}(\mathbf{\hat{p}}) ,
\end{align}
where $N =  \sqrt{\frac{E + m}{2m}}$, $\tilde{p} = \frac{|\mathbf{p}|}{E + m}$, $\mathbf{\hat{p}}$ is a unit vector in the direction of $\mathbf{p}$, and $\chi_{\pm \lambda}(\mathbf{\hat{p}})$ are given in Table \ref{ta:spinors}.

\begin{table}
\caption{Two component spinors of (\ref{eq:spinors})} 
\begin{tabular}{|c|c c} 
                &  $\chi_{\frac{1}{2}}(\mathbf{\hat{p}})$ &  $\chi_{-\frac{1}{2}}(\mathbf{\hat{p}})$  \\ \hline \hline \\
initial state   &   $ \left( \begin{array}{c} 1 \\ 0 \end{array} \right)$    & $ \left( \begin{array}{c} 0 \\ 1 \end{array} \right) $   \\ \\ \hline   \\  
final state	&   $ \left( \begin{array}{c} \cos{\frac{\theta}{2}} \\  \sin{\frac{\theta}{2}} \end{array} \right)   $ 
     & $ \left( \begin{array}{c}    -\sin{\frac{\theta}{2}}   \\ \cos{\frac{\theta}{2}}    \end{array} \right)  $
\end{tabular}
\label{ta:spinors}
\end{table}
\section{Amplitudes to Fermi Invariants}
The helicity dependent matrices which relate the Fermi invariants to the helicity amplitudes are,
\begin{equation}
 C_{ij}^{t} = 
\left( \begin{array}{ccccc}
1+\frac{t}{s-4m^2}      &    -1 + \frac{s}{2m^2} +\frac{t}{s-4m^2}             & -2 + \frac{2t}{s-4m^2} & 0  & -1 + \frac{s}{2m^2} - \frac{t}{s-4m^2} \\ \\
C^{t}_{21}              & C^{t}_{21}                                           & 2C^{t}_{21}            & 0  &-C^{t}_{21}                             \\ \\
1+\frac{t}{s-4m^2}      &  C^{t}_{32} & 2+\frac{2t}{s-4m^2}   & 0  & -C^{t}_{32}  \\ \\
\frac{st}{4m^2(s-4m^2)} & \frac{t}{s-4m^2}   & \frac{s-2m^2}{m^2}\left( 2+\frac{t}{s-4m^2} \right)      & \frac{t}{4m^2} &  -2-\frac{t}{s-4m^2}         \\ \\
\frac{-st}{4m^2(s-4m^2)}& \frac{-t}{s-4m^2}  &  \frac{-2t}{s-4m^2}                         & \frac{t}{4m^2}           & \frac{t}{s-4m^2} \\ \\
\end{array} \right)
\end{equation}

\begin{equation}
C^{t}_{21}=-\frac{\sqrt{s}}{4m}\sin(\theta) =  -\frac{\sqrt{s}}{2m}\sqrt{\frac{-t}{s-4m^2}} +\frac{\sqrt{s}}{4m}\left(\frac{-t}{s-4m^2} \right)^\frac{3}{2}
\end{equation}
\begin{equation}
C^{t}_{32}= \frac{1}{2m^2}(s-2m^2)\left(1+\frac{t}{s-4m^2}\right)
\end{equation}
\begin{align}
 C_{ij}^{u} = 
\left( \begin{array}{ccccc}
-1-\frac{u}{s-4m^2}     & 1 - \frac{s}{2m^2} -\frac{u}{s-4m^2}             & 2 - \frac{2u}{s-4m^2} & 0  & 1 - \frac{s}{2m^2} + \frac{u}{s-4m^2} \\ \\
-\sqrt{\frac{(4m^2-s-u)su}{4m^2(s-4m^2)^2}} &-\sqrt{\frac{(4m^2-s-u)su}{4m^2(s-4m^2)^2}}&-2\sqrt{\frac{(4m^2-s-u)su}{4m^2(s-4m^2)^2}}&0& \sqrt{\frac{(4m^2-s-u)su}{4m^2(s-4m^2)^2}} \\ \\
\frac{-su}{4m^2(s-4m^2)} & \frac{-u}{s-4m^2}  & \frac{-2u}{s-4m^2}  & \frac{u}{4m^2}  & \frac{u}{s-4m^2}  \\ \\
\frac{-su}{4m^2(s-4m^2)} & \frac{-u}{s-4m^2}   & \frac{s-2m^2}{m^2}\left(-2-\frac{u}{s-4m^2} \right)   & -\frac{u}{4m^2} & 2+\frac{u}{s-4m^2} \\ \\
1+\frac{u}{s-4m^2}       & \frac{s-2m^2}{m^2}\left(-1-\frac{u}{s-4m^2} \right)  &  2+\frac{2u}{s-4m^2} & 0 & \frac{s-2m^2}{m^2}\left(-1-\frac{u}{s-4m^2} \right) \\ \\
\end{array} \right)
\end{align}
\section{Electromagnetic Effects}
We utilize the full proton vertex given as,
\begin{align} \label{eq:EM_vertex}
\Gamma^{\mu}_{EM} = F_1(t) \gamma^{\mu} - \frac{F_2(t)}{2m} & i \sigma^{\mu \nu}q_{\nu}, \\
  F_1(t) = \frac{G_E(t) - G_M(t) t/4m^2}{1 - t/4m^2} &\quad F_2(t) = \frac{G_M(t) - G_E(t)}{1 - t/4m^2} \\
    G_E = G_M/2.79 =& (1 - t/.71)^{-2},
\end{align}
where $q_{\nu}$ is the four momentum of the photon. 

The one photon-exchange contribution to the helicity amplitudes is then given as,
\begin{align}\label{eq:EMamp}
a_{EM}(s,t) &=  \frac{4\pi/137}{2t(4m^2 - s)(-4m^3 + mt)^2}\times \\   
       &(-8G_{E}G_{M}{m^2}stu- G_{M}^2t(32m^6 + s^2t + 2m^2t(s + t) - 8m^4(s + 2t))   \\
       &- 8G_{E}^2m^4(16m^4 + 2s^2 + 3st + t^2 - 4m^2(3s + 2t)))  \\
b_{EM}(s,t) &= -\frac{4\pi/137}{2mt(-4m^2 + t)^2}\sqrt{\frac{stu}{(s - 4m^2)^2} }\times \\
       &\left( (s-u)(4m^2G_{E}^2 + G_{M}^2 t) + 2G_{E}G_{M}(16m^4 - st - 4m^2(s + t) )   \right)  \\
c_{EM}(s,t) &= -\frac{(4\pi/137)u}{2t(4m^2 - s)(-4m^3 + mt)^2}\times \\
       &(8G_{E}^2m^4(u-s) + 8G_{E}G_{M}m^2st + G_{M}^2t(-8m^4 + 2m^2t - st))\\
d_{EM}(s,t) &=  \frac{4\pi/137}{(s - 4m^2)(t-4m^2)^2} \times\\
       &(4G_{E}G_{M}su + G_{E}^2s(s-u) + G_{M}^2(16m^4 + 2s^2 + 3st + t^2 - 4m^2(3s + 2t))) \\
e_{EM}(s,t) &=  -\frac{4\pi/137}{(s-4m^2)(t-4m^2)^2}\times \\
	&(4G_{E}G_{M}su + G_{E}^2s(s-u) + G_{M}^2(16m^4 + 2s^2 + 3st + t^2 - 4m^2(3s + 2t))).
\end{align}
where $u = 4m^2 -s -t$.
\begin{acknowledgments}
We thank the Jefferson Science Associates for providing the JSA/JLab Graduate Fellowship to WF. We also thank J. Dudek and A. Prokudin for the helpful discussions.
\end{acknowledgments}

\bibliography{Reggemodel.bib}

\end{document}